# Effective Caching for the Secure Content Distribution in Information-Centric Networking

Muhammad Bilal[1], Shin-Gak Kang[2] and Sangheon Pack[3]

[1] Hankuk University of Foreign Studies, Yongin-si, Rep. of Korea
[2] Electronics and Telecommunications Research Institute, Daejeon, Rep. of Korea
[3] Korea University, Seoul, Rep. of Korea
emails: m.bilal@ieee.org, sgkang@etri.re.kr, shpack@korea.ac.kr

*Abstract*—**The secure distribution of protected content requires consumer authentication and involves the conventional method of end-to-end encryption. However, in information-centric networking (ICN) the end-to-end encryption makes the content caching ineffective since encrypted content stored in a cache is useless for any consumer except those who know the encryption key. For effective caching of encrypted content in ICN, we propose a novel scheme, called the Secure Distribution of Protected Content (SDPC). SDPC ensures that only authenticated consumers can access the content. The SDPC is a lightweight authentication and key distribution protocol; it allows consumer nodes to verify the originality of the published article by using a symmetric key encryption. The security of the SDPC was proved with BAN logic and Scyther tool verification.**

## I. Introduction

SINCE the earliest time of the Internet, its underlying architecture has been based on packet-switching and host-to-host communication. The TCP/IP layered architecture employs the same view and provides an abstract host-to-host communication model to communication applications. However, in the recent past there has been a profound increase in Internet connectivity, and with the emergence of new Internet applications, the Internet semantics have changed from host centric to content centric. To satisfy the needs of emerging internet applications, the current TCP/IP Internet architecture has adopted several application layer solutions known as Over-the-Top (OTT) applications, such as Content Delivery Network (CDN), web caching, and peer-to-peer networking [1-3]. The additions of new OTT applications are leading us towards a very complex internet architecture, and are introducing challenges to achieving efficiency, security, privacy etc.at acceptable economical cost. In this perspective, Information-Centric Networking (ICN) has emerged as a promising architecture for the Future Internet.

ICN represents a paradigm shift from host-centric to content-centric services and from a Source-driven to Receiver-driven approach. In the ICN paradigm the network is then in charge of doing the mapping between the requested content and where it can be found. To do so, network level naming is used for identifying content objects, independent of their location or container [4]. This means that the ICN architecture decouples content from the host at the network level and supports the temporary storage of content in an in-network cache [5-6]. The benefits of the ubiquitous caching in ICN are profound, but it also introduces a challenge to content security.

In an earlier work [7], the author presented a scheme for protected content using network coding as encryption. However, that scheme requires a private connection between the publisher and consumer to obtain the decoding matrix and some missing data blocks. In another study [8], the author presented a security framework for the distribution of encrypted copyright video streaming in ICN. However, each video was encrypted with a large number of symmetric encryptions keys, such that each video frame was encrypted with a unique symmetric encryption key. Only authorized users who possessed the set of all keys could decrypt the video content. The distribution of a large number of keys for each video content is an extra communication overhead.

### A. Problem statement

The distribution of protected content requires the authentication of the consumer and involves a conventional type of end to end encryption. However, in information-centric networking (ICN) the end-to-end encryption for each authorized subscriber makes the content caching ineffective.

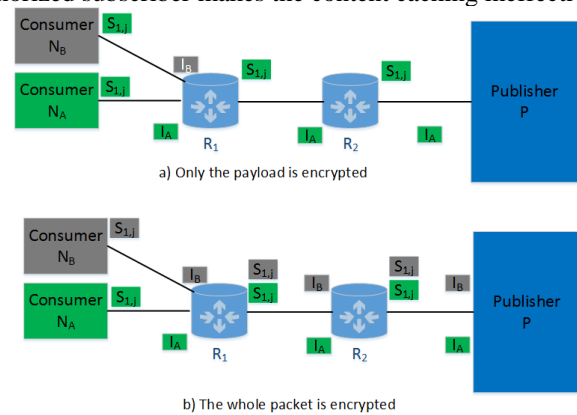

a) Only the payload is encrypted

b) The whole packet is encrypted

**Figure 1 The ineffective caching in ICN with end-to-end encryption**

Muhammad Bilal is supported by the Hankuk University of Foreign Studies research funds for 2018.
Sangheon Pack is supported by R&D program of MOTIE/KEIT [10051306, Development of Vehicular Cloud-based Dynamic Security Framework for Internet of Vehicles (IoV) Services].

Shin-Gak Kang is supported by ICT R&D program of MSIP/IITP. 2016-0-00192, Standards development for service control and contents delivery for smart signage services.

As shown in Figure 1, the consumers $N_A$ and $N_B$ subscribe to the protected content. Consumer $N_A$ sent an interest packet $I_A$ encapsulating authorization information for the content. In reply, based on the subscription information, the publisher $P$ authenticates the consumer and checks the authorization of content object $O_j$ for the consumer $N_A$. If $N_A$ is a valid subscriber than publisher $P$ encrypts the requested content segment $S_{1,j}$ and sends it to consumer $N_A$, encrypted with a consumer specific key. Based on the basic semantics of the information-centric networking (ICN), the intermediate cache routers $R_1$ and $R_2$ stores the encrypted content segment $S_{1,j}$, for future use.

In the next step, consumer $N_B$ requests the same content. As shown in Figure 1-a, if the meta data of the encrypted stored packet is available to $R_1$, then based on the basic semantics of ICN the intermediate cache router $R_1$ will reply with the cached content $S_{1,j}$ to consumer $N_B$. However, consumer $N_B$ cannot decrypt the content segment $S_{1,j}$ as it was solely intended for consumer $N_A$.

On the other hand, as shown in Figure 1-b, if the meta data of the encrypted stored packet is unavailable to $R_1$, that is, the meta data is also encrypted, then the interest packet $I_B$ will be forwarded to the publisher. If $N_A$ is a valid subscriber then publisher $P$ will encrypt the requested content segment $S_{1,j}$ and send it to consumer $N_A$, encrypted with a consumer specific key.

The solution is to encrypt each content segment with a key known to all subscribers; which raises three fundamental questions. How does one ensure that only an authenticated subscribed consumer can access the content? How can the consumer verify the originality of the published article; that is, do we still need self-certifying? Finally, and most importantly, how can encryption keys be distributed among all of the consumers for each content segment? We will answer all these questions in this work.

The remainder of this paper is organized as follows. In Section II, we present a brief system model overview. Section-III describes the proposed scheme with a detailed discussion. In Section-IV, we assess the strength of using BAN logic and Scyther verification. Finally, we provide concluding remarks in Section –V.

## II. System Model and Sketch of Proposed Scheme

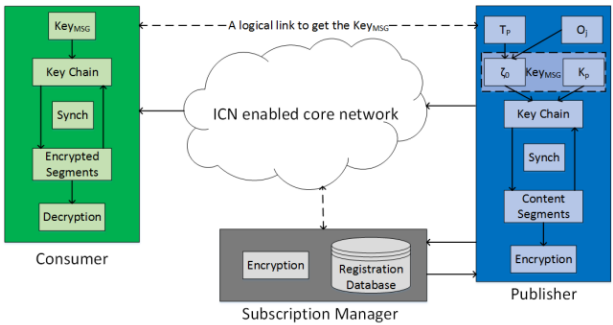

**Figure 2 Illustration of system model used in proposed scheme.**

The system model used throughout this work is shown in Figure 2. To enable the ICN core network to support the effective caching of encrypted content we introduce a new entity, designated subscription manager $M$. We assume that there is a secret number $n_S^i$ associated with each valid subscriber consumer that is known to the subscription manager $M$, that is, the valid consumers are already registered with the subscription manager $M$. Note that being registered doesn't mean the consumer is entitled to access certain protected content. Moreover, subscription manager $M$ can be a module installed on the publisher or it could be an independent entity in the network.

In this work we assume that subscription manager $M$ is an independent entity associated with multiple publishers. This design reduces the message exchange complexity for the case when a consumer decides to subscribe to multiple protected contents published by different publishers.

When a registered consumer is interested in protected content they first need subscribe to the protected content, for instance, subscribing to a movie channel. In the first step, the consumer sends an interest request for the protected content along with the subscription request, and the publisher node routes the request towards subscription manager $M$. The subscription manager $M$ authenticates consumer $N_A$ and in response publisher $P$ sends the encryption key generation information $Key_{MSG}$. Using $Key_{MSG}$ as a seed for a simple hash function, consumer $N_A$ and publisher $P$ can generate a chain of keys. Publisher $P$ uses these keys to encrypt the segments of the published content; likewise, after acquiring $Key_{MSG}$ consumer $N_A$ generates the same keys to decrypt the segments of the published content. Detailed descriptions of the key generation and secure subscription are discussed in subsequent sections.

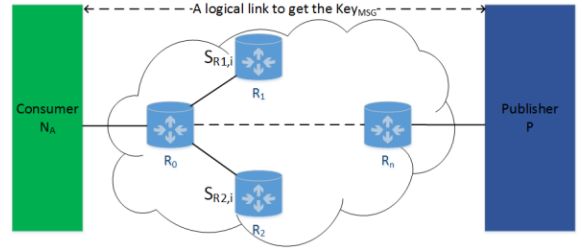

**Figure 3 An example of secure delivery of protected content**

Let's consider the case where consumer $N_A$ is already a registered consumer node. As shown in Figure 3, consumer $N_A$ sends an interest packet $I_{A,i}$ encapsulating authorization information for the protected content object $O_i$. Let us say that protected content object $O_i$ is composed of $k$ number of segments $S = \{S_{1,i}, S_{2,i} \dots S_{k,i}\}$; further, the intermediate cache routers $R_1$ and $R_2$ have copies of the protected content segments, represented by $S_{R1,i} \subseteq S$ and $S_{R2,i} \subseteq S$. If $N_A$ is a valid subscriber then publisher $P$ sends the encryption key generation information $Key_{MSG}$ to consumer $N_A$. After receiving the key generation information, the consumer can decrypt the content segments, which may be delivered directly from the intermediate cache router.

## III. Proposed Scheme

The SDPC protocol suite consists of two protocol suites, the Keying Protocol suite and the Subscription and Content Access Protocol suite. The Keying Protocol suite is comprised of a key generation protocol and a key agreement protocol for content protection. Likewise, the Content Access Protocol further comprises four protocols, one dealing with the consumer node subscription and the other three dealing with access to the protected contents published by the different types of publishers. The SDPC protocol suite is described in detail in subsequent sections.

### A. Keying Protocol Suite

In the keying protocol suite, the key generation protocol generates a 'commitment key' using an irreversible function similar to the ones used in [9-10]. The 'commitment key' is further used to drive multiple keys.

The key generation mechanism for the content protection is shown in Figure 4 and consists of the following steps: 1) The publisher divides the large contents into equal sized segments. 2) For each protected content object $O_j$ the publisher generates a unique commitment key generator by using an irreversible one way hash function $\zeta_0^j = H(T_P, O_j)$, where $T_P$ is the time of publishing and $O_j$ represents the content name and version. 3) The publisher now generate a "Chain of Key Generators" of length $L = \frac{sizeof(o_j)}{segment\ size}$ by using an irreversible one-way function: $\{H(\zeta_0^j) = \zeta_1^j, H(\zeta_1^j) = \zeta_2^j \dots H(\zeta_{l-1}^j) = \zeta_l^j\}$; i-e $H(\zeta_k^j)^i = \zeta_{k+i}^j$. 4) Each generator ($\zeta$) in the chain is used by function $g$ at a specific index location in the chain to derive a content segment encryption key. For instance, at index $k$ the function $g(\zeta_k^j) = H(\zeta_k^j, K_p)$ generates the key $K_k^j$ used for encrypting the $kth$ segment of the content object $O_j$, where $K_p$ is the public key of the publisher.

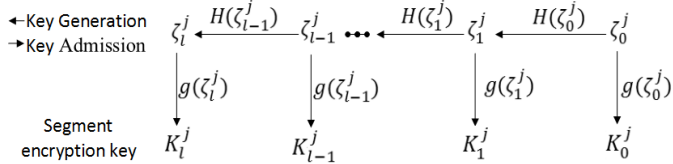

**Figure 4 Symmetric keys generation and admission with reference to segment number of protected content**

The symmetric keys generated as a result of SDPC keying protocol have size of 256 bits (32 bytes); hence, in the subsequent section of authentication protocols any symmetric encryption can be used, e.g., RC5/6; Rijndael, Twofish, MARS, and Blowfish symmetric encryption algorithms support the 256-bit encryption key.

### B. Subscription and Content Access Protocol suite

When a consumer wants to subscribe to the protected content, for instance a movie database, they gain initial access using a subscription protocol (SubP). After SubP the consumer can use the ticket to access multiple protected contents published by the publishers, or managed by a third party. In subsequent sections, the Subscription and Content Access Protocol suite are described in detail.

#### a) Initial Access and Subscription Protocol (SubP)

If a consumer node $N_i$ wants to subscribe to the protected content, for instance subscribing for the movie channel, in the first step, $N_i$ generates an encryption key $K_{TS}^i = H(K_p^j \oplus n_S^i)$, where $K_p^j$ is the public key of publisher and $n_S^i$ is a secret number shared with the subscription manager $M$. The consumer sends an interest request for the protected content along with the subscription request, encrypting the secret number $n_S^i$ with $K_{TS}^i$. The publisher node routes the request towards $M$, and the protocol continues as follows:

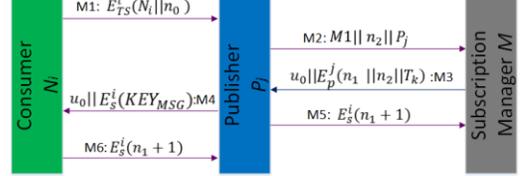

**Figure 5 Message exchange for initial access and subscription protocol**

M1. As shown in Figure 5, $N_i$ injects a subscription interest packet $I_i$. The ICN core network forwards it to the publisher $P_j$. The interest packet encloses the $n_0^i$ which is encrypted with the generated encryption key $K_{TS}^i$.

M2. Upon receiving the request from $N_i$, the $P_j$ forwards the request in conjunction with its identity and challenge $n_2$ to the subscription manager $M$. Note that $P_j$ cannot decrypt the part of the interest packet, which is encrypted with key $K_{TS}^i$ and holds the secret registration number $n_S^i$.

M3. $M$ retrieves the profile from the database, and if $N_i$ is a legitimate consumer, $M$ generates the keys $K_{TS}^i = H(K_p^j \oplus n_S^i)$, $K_S^i = H(T_M \oplus n_S^i)$, and sends $u_0 = E_{TS}^i(n_S^i + 1 \|n_1\|T_k\| K_S^i)$ to $P_j$ in M3, $T_M$ is the time of issuing the session key $K_S^i$. M3 also includes ticket $T_k = E_P^j(N_i\|K_S^i\| profile)$, $n_1$ (a challenge for $N_i$), and $n_2$ (challenge response for the $P_j$), all encrypted with $K_p^j$. The publisher $P_j$ verifies the challenge $n_2$, stores $n_1$ and retrieves the profile and $K_S^i$ from the ticket. Note that the ticket is encrypted with the public key of the publisher. The consumer node $N_i$ cannot decrypt it, but can use it to subscribe to other contents published by the publisher $P_j$, without contacting subscription manager $M$.

M4. $P_j$ forwards the $u_0$ to $N_i$ along with the $KEY_{MSG} = (\zeta_0^j, K_p)$, which is required to decrypt the segments of the published content. After a challenge ($n_0^i + 1$) verification, $N_i$ accepts $T_k$ and generates the key chain to decrypt the protected published content. The key chain is generated using the public key of $P_j$, hence, the content is also self-certifying.

M5. $P_j$ sends the challenge response ($n_1 + 1$) to $M$ for the confirmation of a successful protocol run.

M6. After challenge ($n_0 + 1$) confirmation, $P_j$ may optionally register the $N_i$ in its own database. If $P_j$ does not receive a challenge response within a certain period of time, then $P_j$ marks $T_k$ as a stolen ticket.

In the SubP, a secure exchange of $n_0$ ensures the message authentication between the consumer and the publisher, $n_2$ between the publisher and subscription manger, and $n_1$ between the subscription manger and the publisher, while message authentication between the consumer and publisher is established by session key encryption and $n_1$.

### b) Content Access Protocols

#### (1) Access Protocol after Subscription (APSub)

Further, if consumer $N_i$ wishes to access some other protected contents published by the publisher $P_j$, then $N_i$ sends an interest request for the protected content along with the ticket $T_k$ and the protocol continues as follows:

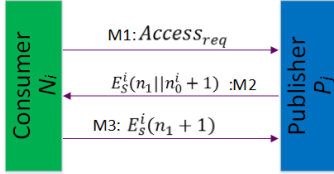

**Figure 6 Message exchange for access protocol after subscription**

M1. As shown in Figure 6, $N_i$ injects a subscription interest packet, enclosing $Access_{req} = E_S^i(N_i \,||n_0)||T_k$. The ICN core network forwards it to the publisher $P_j$. The publisher $P_j$ decrypts the ticket, retrieves $K_s^i$, verifies sender identity $N_i$. If the value $N_i$ does not match, the $P_j$ will ignore the request and otherwise proceed as follows.

M2. $P_j$ sends a challenge response along with the new challenge encrypted with session key $K_s^i$. $P_j$ also send the $Key_{MSG}$, which is required to decrypt the segments of the published content.

M3. $N_i$ sends the challenge response $n_1$. If $P_j$ does not receive the challenge response within a certain period of time, then $P_j$ marks $T_k$ as a stolen ticket.

In the above APSub, the secure exchange of $n_0^i$ ensures the message authentication between the consumer and the publisher.

#### (2) Access Protocol after Subscription involving a Third party (APSub3)

Assume a consumer $N_i$ subscribed with $P_i$, which means it shares a session key $K_s^i$ with $P_i$ and holds a $T_k$ encrypted with public key of $P_i$. Now if $N_i$ wishes to access the protected contents published by a third-party content publisher $P_j$, APSub3 continues as follows:

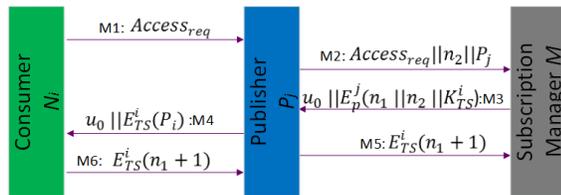

**Figure 7 Message exchange for access protocol after subscription involving a third party**

M1. As depicted in (3) at Figure 3-b, $N_i$ injects a subscription interest packet enclosing $Access_{req} = E_S^i(N_i \,||n_0)||T_k$ and the packet is forwarded to the publisher $P_j$.

M2. Upon receiving the request from $N_i$, $P_j$ forwards the request in conjunction with its identity and the challenge $n_2$ to $M$. Note that $P_j$ cannot decrypt $Access_{req}$ in the interest packet, which is encrypted with a shared session key $K_s^i$ between $N_i$ and $P_i$, which ensures the third-party content distributor cannot misuse the consumer secure information, such as profile and secret share number etc.

M3. $M$ retrieves the profile from $T_k$, and if $N_i$ is a legitimate consumer, $M$ generates the key $K_{TS}^i = H(K_s^i \oplus n_0)$, and sends $u_0 = E_S^i(n_0 + 1\,||n_1\,||Key_{MSG})$ to $P_j$. The message M3 also includes the key $K_{TS}^i$, $n_1$ a challenge for $N_i$, and $n_2$ the challenge response for $P_j$, which are encrypted with public key pf $P_j$. After that, the publisher $P_j$ verifies the challenge response $n_2$ and stores $n_1$. Note that the ticket is encrypted with the public key of $P_i$. Therefore, $N_i$ and third-party publisher $P_j$ cannot decrypt it. Also, $Key_{MSG}$ is inaccessible to $P_i$, which ensures that the third-party content distributor cannot misuse the protected content.

M4. $P_j$ forwards $u_0 \,||E_{TS}^i(P_j)$ to $N_i$. After the verification of the challenge $(n_0 + 1)$, $N_i$ generates $K_{TS}^i = H(K_s^i \oplus n_0)$ and sends the challenge response $(n_1 + 1)$ to $P_j$. Now $N_i$ can generate a key chain to decrypt the protected published content. Since the key chain is generated using the public key of $P_i$, the content is also self-certifying.

M5. $P_j$ sends the challenge response $(n_1 + 1)$ to $M$ for the confirmation of a successful protocol run.

M6. After the challenge confirmation, $P_j$ closes the protocol run. If $P_j$ does not receive any challenge response within a certain period of time, $P_j$ marks $T_k$ as a stolen ticket.

In SubP3, secure exchanges of $n_0$, $n_1$, and $n_2$ ensure the message authentication between the consumer and the subscription manger, between the subscription manger and the third-party publisher, and between the third-party publisher and subscription manger, respectively. On the other hand, the message authentication between the consumer and the third-party publisher is established by a temporary session key $K_{TS}^i$ and $n_1$.

### C. MPEG Video Distribution: An application of the proposed scheme

This section briefly explains how our proposed scheme can be used for the effective distribution of protected MPEG video in ICN. In MPEG the video is defined as a stream of a group of pictures (GOPs). As shown in Figure 8, each GOP consists of one I frame (Intra-coded picture) and multiple P (Predicted picture) and B (Bidirectional predicted picture) frames. To recover the video in its real quality most of the information is stored in the I-Frame. If a publisher $P_i$ publishes a protected MPEG video, by encrypting the I-Frame the video remains protected. The partial encryption of each GOP is the same as the method employed in [7] but with SDPC the subscriber can generate a large number of keys with the exchange of just a single commitment key.

Using our proposed scheme, the publisher $P_i$ generates a chain of keys for the protected content object $O_j$ and encrypts the I-Frame in each GOP using a corresponding key from the key chain; for instance the I-frame of GOP1 is encrypted with $K_1^j$. When a consumer $N_i$ injects the first interest packet, whether enclosing a subscription or access request, the publisher $P_j$ sends the $Key_{MSG}$ to consumer $N_i$. Meanwhile, the intermediate custodian nodes transfer the data to consumer $N_i$. The consumer $N_i$ then generates the chain of keys and decrypts all of the I-Frame segments using the corresponding keys.

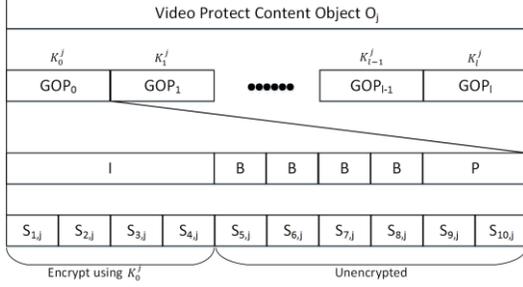

**Figure 8 The structure of video content and usage of SDPC keying protocol.**

## IV. SECURITY ANALYSIS

This section presents an inclusive security analysis of the SDPC protocol using BAN logic [11], and also presents the Scyther [12] implementation result of the SDPC.

### A. Formal security analysis using BAN logic

BAN logic [11] is widely used for the formal analysis of security protocols. To verify the security of the SDPC protocol suite it is sufficient to demonstrate the security of the SubP protocol; the rest of the protocols are extensions of the SubP, which use the ticket and session key established in the SubP protocol run. The three basic objects of BAN logic are principals, formula/statements, and encryption keys. The principals, the protocol participants, are represented by symbols $P$ and $Q$. The formula/statements are symbolized by $X$ and $Y$ and represents the content of the message exchanged. The encryption keys are symbolized by $K$. The logical notations of BAN-logic used for our analysis is given below:

- $P \vDash X$: $P$ believes $X$, or $P$ would be enabled to believe $X$; in conclusion, $P$ can take $X$ as true.
- $P \lessdot X$: $P$ sees/receives $X$. $P$ initially has or received a message $X$ and $P$ can see the contents of the message and is capable of repeating $X$.
- $P|\sim X$: $P$ once said $X$. $P$ has sent a message including the statement $X$. However, the freshness of message is unknown.
- $P \Rightarrow X$: $P$ controls $X$ and should be trusted for formula/statement $X$.
- $\#(X)$: $X$ is fresh; it says, $X$ never sent by any principal before.
- $P \xleftrightarrow{K} Q$: $P$ and $Q$ shares a key $K$ to communicate in a secure way and $K$ is only known to $P$, $Q$ and a trusted principal.
- $(X)_K$: The statement $X$ is encrypted by key $K$.
- $\{X\}_Y$: It stand for $X$ combined with $Y$. $Y$ is anticipated to be secret and its implicit or explicit presence proves the identity of a principal who completes the $\{X\}_Y$.

Some primary BAN-logic postulates used in the analysis of the SDPC are given below:

- Message meaning rules: $\dfrac{P \vDash P \xleftrightarrow{K} Q, P \lessdot (X)_K}{P \vDash Q|\sim X}$ , $\dfrac{P \vDash P \xleftrightarrow{Y} Q, P \lessdot \{X\}_Y}{P \vDash Q|\sim X}$

- Nonce verification rule: $\dfrac{P \vDash \#(X), P \vDash Q|\sim X}{P \vDash Q \vDash X}$

- Jurisdiction rule: $\dfrac{P \vDash Q \Rightarrow X, P \vDash Q \vDash X}{P \vDash X}$

- Freshness rule: $\dfrac{P \vDash \#(X)}{P \vDash (X,Y)}$

- Believe rule: $\dfrac{P \vDash Q \vDash (X,Y)}{P \vDash X, P \vDash Y}$

- Session key rule: $\dfrac{P \vDash Q \#(X), P \vDash Q \vDash X}{P \vDash P \xleftrightarrow{K} Q}$

#### a) Ban Logic Analysis of SubP:

The *SubP* protocol should achieve the following goals which states that both the consumer and the publisher trust the encryption key $K_S^i$ for the secure exchange of $KEY_{MSG}$ :

1. $N_i \vDash (N_i \xleftrightarrow{K_S^i} P_j)$
2. $N_i \vDash P_j \vDash (N_i \xleftrightarrow{K_S^i} P_j)$
3. $P_j \vDash (N_i \xleftrightarrow{K_S^i} P_j)$
4. $P_j \vDash N_i \vDash (N_i(N_i \xleftrightarrow{K_S^i} P_j)$

Protocol Idealization:

M1. $N_i \xleftrightarrow{Via\ P_j} M : \left\{ n_0, \left( N_i \xleftrightarrow{n_S^i} M \right) \right\}_{H(X_S)}$

M2. $P_j \to M : \{n_2,\ IDP_j\}$

M3. $M \to P_j : \left\{ \left( n_1, n_2, \left( N_i \xleftrightarrow{K_S^i} P_j, \#\left( N_i \xleftrightarrow{K_S^i} P_j \right) \right)_{K_P^i} \right)_{K_P^i}, \left\{ n_0, n_1, \left( N_i \xleftrightarrow{K_S^i} P_j, \#\left( N_i \xleftrightarrow{K_S^i} P_j \right) \right)_{K_S^i}, N_i \xleftrightarrow{K_S^i} P_j, \#\left( N_i \xleftrightarrow{K_S^i} P_j \right), N_i \xleftrightarrow{n_S^i} M \right\}_{H(X_S)} \right\}$

M4. $P_j \xleftrightarrow{Via\ M} N_i : \left\{ n_0, n_1, \left( N_i \xleftrightarrow{K_S^i} P_j, \#\left( N_i \xleftrightarrow{K_S^i} P_j \right) \right)_{K_P^i}, N_i \xleftrightarrow{K_S^i} P_j, \#\left( N_i \xleftrightarrow{K_S^i} P_j \right), N_i \xleftrightarrow{n_S^i} M \right\}_{H(X_S)}$

M5. $P_j \to N_i : (KEY_{MSG})_{K_S^i}$

M6. $N_i \to P_j : (n_1)_{K_S^i}$

Initial State Assumptions:

A1. $M \vDash \#(n_0)$
A2. $M \vDash \#(n_2)$
A3. $P_j \vDash \#(n_1)$
A4. $N_i \vDash \#(n_1)$
A5. $N_i \vDash \left( N_i \xleftrightarrow{K_{TS}=H(X_S)} M \right)$
A6. $M \vDash \left( N_i \xleftrightarrow{K_{TS}=H(X_S)} M \right)$
A7. $P_j \vDash \left( P_j \xleftrightarrow{K_P^i} M \right)$
A8. $M \vDash \left( P_j \xleftrightarrow{K_P^i} M \right)$

A9. $M \models N_i \models \left( N_i \xrightarrow{K_{TS}=H(X_S)} M \right)$

A10. $N_i \models M \models \left( N_i \xrightarrow{K_{TS}=H(X_S)} M \right)$

A11. $M \models P_j \models \left( P_j \xleftrightarrow{K_P^j} M \right)$

A12. $P_j \models M \models \left( P_j \xleftrightarrow{K_P^j} M \right)$

Let us analyze the protocol to show that $N_i$ and $P_j$ share a session key:

From M1, we have

$M \vartriangleleft \left\{ n_0, \left( N_i \xleftrightarrow{n_S^i} M \right) \right\}_{H(X_S)}$   (1)

The (1), A6 and message meaning rule infers that

$M \models N_i \mid\sim \left\{ n_0, \left( N_i \xleftrightarrow{n_S^i} M \right) \right\}$ (2)

The A1 and freshness conjuncatenation comprehends that

$M \models \# \left\{ n_0, \left( N_i \xleftrightarrow{n_S^i} M \right) \right\}$   (3)

The (2), (3) and nonce verification rule deduces that

$M \models \left\{ N_i \models n_S^i, n_0, \left( N_i \xleftrightarrow{n_S^i} M \right) \right\}$   (4)

The (4) and believe rule infers that

$M \models N_i \models \left( N_i \xleftrightarrow{n_S^i} M \right)$   (5)

From A2, (5) and jurisdiction rule, it concludes

$M \models \left( N_i \xleftrightarrow{n_S^i} M \right)$   (6)

This belief confirms that $M$ has received a message from a legitimate $N_i$.

From M2, we have

$M \vartriangleleft n_2$   (7)

The (7) and message meaning it infers that

$M \models P_j \mid\sim n_2$   (8)

The A2, A1, (3) and freshness conjuncatenation comprehends that

$M \models \# \left\{ n_0, n_2, \left( N_i \xleftrightarrow{n_S^i} M \right) \right\}$   (9)

According to nonce freshness, this proves that $M$ confirms that $N_i$ is recently alive and running the protocol with $M$.

From M3, we have

$P_j \vartriangleleft \left( n_1, n_2, \left( N_i \xleftrightarrow{K_S^i} P_j, \# \left( N_i \xleftrightarrow{K_S^i} P_j \right) \right), n_0^i \right)_{K_P^i}$   (10)

The A7 and (10) deduce that

$P_j \models M \mid\sim \left\{ n_1, n_0^i, N_i \xleftrightarrow{K_S^i} P_j, \# \left( N_i \xleftrightarrow{K_S^i} P_j \right) \right\}$   (11)

The A3, (11) and freshness conjuncatenation comprehends that

$P_j \models \# \left\{ n_1, n_0^i, N_i \xleftrightarrow{K_S^i} P_j, \# \left( N_i \xleftrightarrow{K_S^i} P_j \right) \right\}$   (12)

The (11), (12) and nonce verification rule infers that

$P_j \models M = \left\{ n_1, n_0^i, N_i \xleftrightarrow{K_S^i} P_j, \# \left( N_i \xleftrightarrow{K_S^i} P_j \right), V_i \right\}$   (13)

The (13) and believe rule comprehends that

$P_j \models M = \left( N_i \xleftrightarrow{K_S^i} P_j \right)$   (14)

The logic belief proves that $P_j$ is confident and believes that $K_S^i$ is issued by $M$; moreover, the freshness of the key also suggests that $M$ is alive and running the protocol with $P_j$ and $N_i$.

The (13), (14) and jurisdiction rule concludes that

$P_j \models \left( N_i \xleftrightarrow{K_S^i} P_j \right)$   (15) Goal-3

From M4, we have

$N_i \vartriangleleft \left\{ n_1, N_i \xleftrightarrow{K_S^i} P_j, \# \left( N_i \xleftrightarrow{K_S^i} P_j \right), N_i \xleftrightarrow{n_S^i} M \right\}_{H(X_S)}$   (16)

The (16), A5 and message meaning rule comprehends that

$N_i \models M \mid\sim \left\{ n_1, N_i \xleftrightarrow{K_S^i} P_j, \# \left( N_i \xleftrightarrow{K_S^i} P_j \right) \right\}$   (17)

The (17), A4 and freshness conjuncatenation rule infers that

$N_i \models \# \left\{ n_1, N_i \xleftrightarrow{K_S^i} P_j, \# \left( N_i \xleftrightarrow{K_S^i} P_j \right) \right\}$   (18)

The (17), (18) and nonce verification rule deduce that

$N_i \models M = \left\{ n_1, N_i \xleftrightarrow{K_S^i} P_j, \# \left( N_i \xleftrightarrow{K_S^i} P_j \right) \right\}$   (19)

The (19) and believe rule infers that

$N_i \models M = \left\{ N_i \xleftrightarrow{K_S^i} P_j \right\}$   (20)

The (19), (20) and jurisdiction rule concludes that

$N_i \models \left\{ N_i \xleftrightarrow{K_S^i} P_j \right\}$   (21) Goal-1

From M5, we have

$N_i \vartriangleleft IDS_j$   (22)

The (15), (21), (22) and meaning rule comprehends that

$P_j \models N_i = \left\{ N_i \xleftrightarrow{K_S^i} P_j \right\}$   (23) Goal-4

From M6, we have

$P_j \vartriangleleft n_1$   (24)

The (15), (21), (23) and nonce verification rule deduce that

$N_i \models P_j = \left\{ N_i \xleftrightarrow{K_S^i} P_j \right\}$   (23) Goal-2

### B. Verifying the protocol using the Scyther tool

The previous section proved that according to the BAN logic the SDPC is a secure authentication scheme. The Ban logic provides a foundation for the formal analysis of security protocols, but few attacks can slip through the BAN logic [11]. For further proof of the strength of the SDPC protocol suite, we implemented the SDPC in an automated security protocol analysis tool, Scyther [12]. We considered four claims: Aliveness, weak agreement, non-injective agreement, and non-injective synchronization. For a detailed description of the protocol claims, please refer to [13-14].

Table 5.1 SCYTHER TOOL PARAMETER SETTINGS

| Parameter | Settings |
|---|---|
| Number of Runs | 1~3 |
| Matching Type | Find all Type Flaws |
| Search pruning | Find All Attacks |
| Number of pattern per claim | 10 |

In Scyther the protocol is modeled as an exchange of messages among different participating 'roles'; for instance, the consumer node is in the role of initiator, the publisher is in the role of responder and the subscription manger is in the role of a server. The Scyther tool integrates the authentication properties into the protocol specification as a claim event. We tested our protocol by employing the claims mentioned earlier, with the parameter settings given in Table I.

Figure 9 Scyther results for the SDPC SubP protocol

The protocol is tested using given parameter in Table I. The results are shown in Figure 9. It is clear that SubP protocol qualifies all the protocol claims and no attacks were found. Hence, for a large number of systems and scenarios, our protocol guarantees safety against a large number of known attacks, such as impersonating, man-in-middle    and replay attacks, etc.

## V. Conclusions

In information-centric networking (ICN), end-to-end encryption for each subscriber makes content caching ineffective, since encrypted content stored in a cache is not useful for any other consumer except those consumers who know the encryption key. For effective caching of encrypted content, we proposed a novel scheme, called the Secure Distribution of Protected Content (SDPC). In the SDPC scheme we designed two protocol suites, the Keying Protocol suite and

Subscription and Content Access Protocol suite. The Subscription and Content Access Protocol suite ensures that only authenticated consumers can access the content; hence, providing protection to content. The SDPC's keying protocol suite empowers the publisher and consumer to generate multiple symmetric encryption keys with the exchange of just a single commitment key. The commitment key is generated with the publishers' public key, along with other secret credentials, and allows the consumer to verify the originality of the published article. In other words, self-certifying is achieved with symmetric key encryption. In the conventional ICN architecture, the self-certifying is achieved by means of asymmetric cryptography, which is computationally much more expensive compared to symmetric key encryption. Hence, SDPC is a lightweight and efficient solution for the secure content distribution in ICN.

## References

[1] C. Ge, Z. Sun, and N. Wang, "A Survey of Power-Saving Techniques on Data Centers and Content Delivery Networks," *IEEE Communications Surveys and Tutorials*, vol. 15, no. 3, 3rd Quarter 2013.

[2] S. Podlipnig and L. Boszormenyi, "A Survey of Web Cache Replacement Strategies," *ACM Computing Surveys*, vol. 35, no. 4, Dec. 2003.

[3] A. Malatras, "State-of-the-art survey on P2P overlay networks in pervasive computing environments," *Journal of Network and Computer Applications*, vol. 55, Sept. 2015.

[4] B. Ahlgren, C. Dannewitz, C. Imbrenda, D. Kutscher, and B. Ohlman, "A Survey of Information-Centric Networking," *IEEE Communications Magazine*, vol. 50, no. 7, July 2012.

[5] M. Bilal and S. G. Kang, "Time Aware Least Recent Used (TLRU) Cache Management Policy in ICN," in *Proc.IEEE ICACT'14* Pyeongchang, Korea (South), Feb. 2014.

[6] M. Bilal and S. G. Kang, "A cache management scheme for efficient content eviction and replication in cache networks," *IEEE Access*, vol. 5, pp. 1692-1701, 2017, DOI: 10.1109/ACCESS.2017.2669344.

**[7]** E. Cho, J. Shin, J.Choi, T. Kwon and Y. Choi, "A Tradeoff between Caching Efficiency and Data Protection for Video Services in CCN," in *Proc. Workshop on Security of Emerging Networking Technologies*,San Diego, California, USA, Feb. 2014.

[8] T. Xiaobin, J. Liguo, Z. Zifei and Y. Pei, "Copyright Protection Scheme for Information-Centric Networking Base on the Linear Network Coding," In *Proc. 35th Chinese Control Conference*, Chengdu, China, July 27-29, 2016.

[9] A. Perrig, R. Canetti, J. D. Tygar and D. Song. The TESLA Broadcast Authentication Protocol. Available online: http://www.cs.berkeley.edu/~tygar/papers/TESLA_broadcast_authentication_protocol.pdf, 2002.

[10] M. Bilal and  SG. Kang, "An Authentication Protocol for Future Sensor Networks," *Sensors* 2017, 17, 979.

[11] M. Burrows, R.Needham Abadi, "A logic of authentication," *ACM Trans. Comput.Syst*. 8 (1990) 18–36.

[12] C. Cremers, "The Scyther Tool: Verification, Falsification, and Analysis of Security Protocols," Available online: https://www.cs.ox.ac.uk/people/cas.cremers/downloads/papers/ Cr2008-Scyther_tool.pdf

[13] G. Lowe. A Hierarchy of Authentication Specifications. In proceedings. 10th Computer Security Foundations Workshop. June 1997

[14] C. Cremers and S Mauw and E.P. De Vink. Injective synchronisation: An extension of the authentication hierarchy. Theoretical Computer Science, vol. 367, no. 1-2, November 2006